\documentclass[]{spie}
\usepackage[dvips]{graphicx}

\title{A Distributed Control System for Rapid Astronomical Transient
	Detection}

\author{J. Wren, K. Borozdin, S. Brumby, D. Casperson, M. Galassi, \\
        K. McGowan, D. Starr, W. T. Vestrand, R. White, and P. Wozniak}

\authorinfo{All authors affiliated with Los Alamos National 
	Laboratory, Los Alamos, NM 87545\\Send correspondence to
	J. Wren: E-mail: jwren@lanl.gov, Telephone: (505)665-5941}

\begin{document}
\maketitle

\begin{abstract}
The Rapid Telescope for Optical Response (RAPTOR) program consists
of a network of robotic telescopes dedicated to the search for
fast optical transients.  The pilot project is composed of three
observatories separated by approximately 38 kilometers located near
Los Alamos, New Mexico.  Each of these observatories is composed of a
telescope, mount, enclosure, and weather station, all operating
robotically to perform individual or coordinated transient searches. 
The telescopes employ rapidly slewing mounts capable of slewing a 250
pound load 180 degrees in under 2 seconds with arcsecond precision.
Each telescope consists of wide-field cameras for transient detection
and a narrow-field camera with greater resolution and sensitivity. 
The telescopes work together by employing a closed-loop system for
transient detection and follow-up.  Using the combined data
from simultaneous observations, transient alerts are generated and
distributed via the Internet.  Each RAPTOR telescope also has the
capability of rapidly responding to external transient alerts received
over the Internet from a variety of ground-based and satellite
sources.  Each observatory may be controlled directly, remotely, or
robotically while providing state-of-health and observational results
to the client and the other RAPTOR observatories.  We discuss the
design and implementation of the spatially distributed RAPTOR system.
\end{abstract}


%
\section{INTRODUCTION}
\label{sect:intro}

Constructing an observatory that can run in a completely autonomous
manner is a significant challenge.  Not only must it be capable of
scheduling its own observations, but it must robustly handle things
such as changing weather conditions, software errors, and hardware
failures.  Recently, several projects have met this challenge
admirably.  The recent detection of a prompt optical flash from the
gamma-ray burst GRB990123 by the Robotic Optical Transient Search
Experiment I (ROTSE-I) at Los Alamos National Laboratory demonstrated
the value of small robotic observatories\cite{Akerlof99}.  The 80
second flash from GRB990123 was the brightest object ever observed in
optical wavelengths at an absolute magnitude of -36.

The detection of GRB990123 also demonstrates how poorly the night sky
is monitored for transient objects.  Gamma-ray bursts occur every day,
yet the ROTSE-I detection is still the only contemporaneous optical
detection of a GRB.  Undoubtedly many more events like this have been
within the reach of small telescopes, yet few have been observed.
The aim of the RAPTOR project is to develop a wide-field system for
detecting these optical transients\cite{Vestrand01}.  The system
should have the greatest possible sensitivity for the given field of
view and also be able to operate robotically.  A fast analysis
pipeline is also necessary if transients are to be identified in near
real-time.

The ability to reject false triggers is an absolute necessity for any
program that intends to do a wide-field search for transients that
occur on the timescale of minutes.  This was a significant problem
in previous experiments such as the Explosive Transient Camera
(ETC)\cite{Vanderspek94}.  There can be many causes of false
triggers in any astronomical imaging system; hot pixels, satellite
glints, meteors, camera defects, cosmic rays, etc.  One way to reduce
the number of
false triggers is to require that any object must be present in more
than one image to be considered a valid detection.  This technique can
eliminate many false triggers, however things such as a glint from a
geostationary satellite may still get through.  An additional
technique for eliminating false triggers is to use coordinated
observations from spatially separated observatories.  It is very
unlikely that a camera defect will appear at the same location on two
different telescopes.  Additionally, if the observatories are
separated by a sufficient distance, parallax can be used to eliminate
nearby objects such as satellites and meteors.

Developing a set of telescopes that can operate in a coordinated
manner further aggravates the challenge of building fully robotic
observatories.  Not only must each observatory be able to operate
autonomously, they must also be able to communicate with each other
and synchronize their observations.  Further, if they are going to
search for transients in real-time, they must be able to process and
compare their observations with each other in a fast automated way.
The goal of the RAPTOR project is to build just such a system.

To meet these goals, we have decided on a system similar to human
vision.  Humans use two eyes not just for depth perception, but also to
filter out artifacts that appear in only one eye.  We have a
central fovea which has greater resolution and color sensitivity.  Our
vision process operates in a feedback loop which can quickly
identify a transient object at the edge of the field of view and
``slew'' our eyes to center the object in the foveas to study the
object with greater sensitivity.  Following this example, we have
decided to construct two telescopes, RAPTOR A and B, which will
operate simultaneously.  Each telescope will have a set of four wide
field cameras and a central ``fovea'' camera.  A real-time image
processing pipeline will identify transients in the wide field
system.  If a transient is present in both telescopes at the same
location and the same time, each telescope will re-center the object
in the fovea camera.  Additionally we are constructing a third system,
RAPTOR S, which will consist of a single 12 inch telescope with a
transmission grating allowing low resolution spectroscopic follow-up
of any transients identified by the RAPTOR A-B system.

%
\section{The RAPTOR Observatories}
\label{sect:project}

The RAPTOR A and B telescopes are located at Los Alamos National
Laboratory (LANL) and consist of two identical systems separated by
approximately 38 km.  The RAPTOR A telescope is located at the Fenton Hill
site of LANL at an elevation of $\sim$8700 ft.  The RAPTOR B
telescope is located at the Los Alamos Neutron Science Center (LANSCE)
site which is just southeast of Los Alamos itself at an elevation of
$\sim$7000 ft.  The RAPTOR S telescope is located next to RAPTOR A at
the Fenton Hill site.  The 38 km separation of RAPTOR A and B
generates sufficient parallax to filter out objects in Earth orbit
even out to geostationary altitudes.

\subsection{Cameras and Optics}

The wide-field cameras on the RAPTOR A and B telescopes consist of an
85 mm f1.2 Canon FD lens mounted on an Apogee AP-10 CCD
camera\footnote{Apogee Instruments Inc., www.ccd.com}.  The AP-10
uses a 2000 x 2000 pixel Thompson chip measuring 28 mm on a side.  This
camera/lens combination provides a 19$^{o}$ x 19$^{o}$ field to a
limiting magnitude of 13.  The fovea cameras consist of a 400 mm f2.8 Canon 
FD lens mounted on a Finger Lakes 1000 x 1000 pixel CCD
camera\footnote{Finger Lakes Instrumentation, LLC, www.fli-cam.com}.
The fovea systems should provide a 2$^{o}$ x 2$^{o}$ field of view to
a limiting magnitude nearing 17.  Both of the CCD cameras are
thermoelectrically cooled and provide a fast readout of $\sim$5
seconds.  Focus on all of the cameras is controlled using a 1/4-80
threaded screw which is adjusted by hand.  Temperature variations do
cause focus shifts, but seasonal re-focusing is adequate to maintain
satisfactory performance of the cameras.

The RAPTOR S optical system consists of an Apogee AP-6 CCD camera
mounted on a 12 inch aperture f/7 Richey-Chretien telescope.  The
telescope and camera rest in the exact same mount as the RAPTOR A and
B telescopes.  The telescope itself has motorized focus control;
however it also has invar spacers which should reduce the need to
re-focus as the temperature changes.

Each camera has its own dedicated dual-processor 1 GHz Pentium III
computer running the Red Hat Linux distribution\footnote{Red Hat Inc.,
www.redhat.com}.  The camera computers run a server which accepts
commands over a socket connection.  The resulting images are stored
locally on hard disk until they are transferred to the LANL tape storage
system at a later time.  The network connection at RAPTOR B is fast
enough that we simply transfer processed images over the network to
the tape storage system.  For RAPTOR A and S the network connection is
slow enough that transferring over the network is impractical.  For
those telescopes we visit the site on a regular basis and download the
images to portable hard drives which we then bring back to LANL
headquarters to be loaded on to the tape system.


\subsection{Mounts}

All RAPTOR optical systems are attached to a rapidly slewing mount
custom built for this project by The Pilot Group\footnote{J. Alan
Schier, The Pilot Group, Monrovia CA, (626)599-9422}.
These mounts employ a servo-motor control system using encoders
consisting of a precision ruled tape with 10 micron spacing in close
proximity to opto-electronic sensors that detect relative motion.  This
encoder system can provide arcsecond pointing accuracy relative to the
home position of the mount.  The mounts are capable of slewing at a
speed of nearly 100$^{o}/s$ and accelerating to this velocity in less
than one second.  Emergency braking is accomplished using a
pneumatically driven caliper which clamps on to a thin steel disk
located on each drive axis.  The mounts are controlled by an
independent PC running Windows NT 4.0 which accepts simple text based
commands over the serial port.

When the system begins operations each night, the mount finds its
absolute home position by slewing to the hard limits in both Right
Ascension and Declination.  The absolute home position can vary by
several arcminutes so a single image is taken at the start of the
night and processed to find the position of the image center in sky
coordinates.  The relative mount coordinates are then adjusted to
correspond to the sky coordinates from the test image.  The resulting
accuracy is a few arcseconds, well below our pixel scale.


\subsection{Observatory Enclosures}

The RAPTOR telescopes are housed inside 10 ft diameter welded steel
cylindrical enclosures built for us by D \& R Tank Co. in Albuquerque,
New Mexico (Fig. \ref{fig:enclosure}).  The roof of the enclosure is
covered by an aluminum clamshell attached to a linear actuator.  The
clamshell is large enough that the telescope may be moved freely
underneath it when the clamshell is closed.  When the telescope is ready
for observations after sunset, the actuator drives open the clamshell
providing the telescope with an unobstructed field of view.  The
telescope mount sits on a 30 inch diameter steel pier make of 1/4 inch
thick welded steel.  Both the cylindrical tank and the pier are bolted
to a 10 ft x 12 ft steel truss composed of 10.5 inch thick welded I-beams.

Control of the clamshell is provided by a custom built I/O Box which,
using the SNAP Ethernet I/O control system from Opto 
22\footnote{Opto 22 Inc., www.opto22.com}, accepts commands
over a socket connection (Fig. \ref{fig:iobox}).
The I/O Box accepts commands to turn on AC
power to open or close the clamshell, reads the open and close limit
switches, controls 4 extra AC outlets, and has a 4 channel
analog-to-digital (ADC) converter.  Additionally, the I/O Box has a watchdog
mechanism which will automatically close the clamshell if it has lost
contact with the host computer.  The I/O Box draws its power
from an uninterruptable power supply (UPS) with enough power to close
the clamshell in the event of a power outage.

The weather station is composed of a Davis Weather Monitor
II\footnote{Davis Instruments Inc., www.davisnet.com} which
senses temperature, humidity, dew point, barometric pressure and wind
speed and direction.  Additionally we use a Vaisala
DRD11A\footnote{Vaisala Inc., www.vaisala.com}
precipitation detector which is read out through one of the ADC
channels in the I/O Box.  If any of the weather variables exceed
limits, the control software will issue an alarm and the system will
be put into a standby state with the clamshell closed.

%
\section{The RAPTOR Data Acquisition Software}
\label{sect:daq}

RAPTOR is different from its predecessors in two significant ways.
First, the scientific goals of the project require the operation of
multiple observatories in synchronization with each other.  As stated
in the introduction, this is not a trivial task.  We require the
observatories to be able to image the same location on the sky within
a second of each other.  Simply scheduling coordinated observations
beforehand does not work due to the fact that our system schedules the
observations dynamically.  For instance, observations at one
observatory may be interrupted by a weather alarm while the other
observatory has perfectly good weather.  To maintain synchronization
between the observatories, we have implemented a client/server
mechanism which allows one observatory to act as the master scheduler
and control other ``client'' observatories.  The master will wait
until all the client observatories are ready to execute the next
command, so that none of the clients will ever fall behind the rest.
If a client observatory loses its connection to the master, it must
then return to scheduling its own observations.

The second way in which the RAPTOR system is different from its
predecessors is the need for a real-time analysis pipeline to meet our
scientific goals.  This means that images must be processed and
searched for transient events on the timescale of our typical
exposure, 30 seconds.  If a transient is found by one observatory, the
results must be compared to the data from the other observatory
to reject false positives, hence the need for synchronized
observations.  In this way we establish the feedback loop where the
observatories generate and respond to their own transient alerts.


To facilitate the goals listed above, we are developing a new software
system for running the RAPTOR telescopes.  This data acquisition (DAQ)
software is composed primarily of a set of three UNIX daemons which
schedule the observations and also monitor and control the hardware
(Fig. \ref{fig:daq}).
These daemons run asynchronously and communicate via TCP/IP sockets.
The first daemon, {\sl controld}, represents the brain of the system and
schedules the observations, listens for transient alerts, and provides
commands to the hardware daemons.  The second daemon, {\sl telescoped},
provides the hardware interface to the telescope mount and cameras.
It executes commands from {\sl controld} that it receives over the socket
connection.  The third daemon is {\sl observatoryd} which is responsible
for monitoring the weather and executing commands to open and close
the clamshell.

\subsection{Controld}

The control daemon, {\sl controld}, is the heart of the RAPTOR DAQ system.
{\sl Controld} is capable of running in four different modes: normal,
server, client, and manual.  When running in ``normal'' mode, {\sl controld}
generates commands for the hardware daemons.  The manner in which it
decides which command to execute next is configurable, but generally
consists of a predefined sky-patrol scheme.  This sky-patrol scheme
can be interrupted by transient alerts, weather alarms, and the sun
passing the elevation threshold.  Additionally, {\sl controld} must evaluate
each patrol position for factors such as field elevation and lunar
distance.

To facilitate coordinated operations of separate observatories,
{\sl controld} can operate in both ``server'' mode and ``client'' mode.
To accomplish this, {\sl controld} has a socket port available for a
client mode connection.  If a socket connection is made on this port,
{\sl controld} will relay commands from another {\sl controld} running at a
separate observatory.  If {\sl controld} loses the connection, it will
return to normal mode operation.  To run in server mode, {\sl controld}
simply performs the same tasks that it does in normal mode operation
but then forwards these commands to the clients.  To run in server
mode, {\sl controld} must be properly configured before the DAQ system is
started.

A stand-alone program provides the manual interface by making a socket
connection to {\sl controld}.  Using this program, the user may check the
status of the system at any time.  If the user then enters the command
to put the system into ``manual'' mode, {\sl controld} will stop sending any
automatic commands and will simply relay commands from the user to the
hardware daemons, {\sl telescoped} and {\sl observatoryd}.  Manual mode operation
also overrides client mode commands to {\sl controld}.  Manual mode
commands do not, however, propagate to the {\sl controld} of client
observatories.  

Another stand-alone program provides alert notices to {\sl controld} at the
various observatories.  This program acts as a clearinghouse for alert
notices by accepting multiple connections and forwarding an alert
coming from one location to all the other clients connected to the
alert server.  The alert server also relays alert notices coming from
the gamma-ray burst coordinate network (GCN)\cite{Barthelmy98} to the
various observatories.  This is necessary because RAPTOR-A and
RAPTOR-S are behind the LANL firewall and can not make a GCN connection
themselves.

\subsection{Telescoped}

The telescope daemon, {\sl telescoped}, is responsible for controlling the
telescope mount and cameras.  It was decided to combine these two
hardware elements due to the fact that they are highly interdependent.
Having a single daemon control both elements allows easier management
of timing issues such as preventing an exposure from starting while
the mount is still in motion.  Also mount status information such as
encoder and sky position are more easily incorporated into the fits
headers of the images.

{\sl Telescoped} has a socket port that will accept a connection from
{\sl controld} over which commands are transmitted.  {\sl Telescoped} accepts
commands to sync, slew, park and halt the mount.  It also accepts
commands to sync and move the focus if the telescope has that
capability.  To operate the CCD it accepts commands to take dark and
light exposures and also abort an exposure.  Finally, it accepts
a general query command and responds with the current status of the
telescope.


Communication to the mount is made via a serial connection to the
dedicated mount computer.  The mount computer runs custom software
from the Pilot Group which, in addition to handling the communication 
with our DAQ system, also monitors the mount status such as limit
switches, encoders, and so forth.  The mount computer only accepts
commands to move to raw encoder positions.  The task of converting
these to sky positions falls on our software.  To do this we have
implemented the TPoint\footnote{Patrick Wallace, TPoint Software,
www.tpsoft.demon.co.uk} software system to generate a pointing model.
This only needs to be done when the telescope is initially installed
or the telescope base has been physically moved.  After the model is
created, the only necessary parameter for transforming sky positions
to encoder positions is the current time.

Communication to the cameras is accomplished via another client/server
mechanism.  Each CCD camera has its own dedicated computer which runs
a small server program that accepts commands for that camera.  When
{\sl telescoped} receives a command from {\sl controld} to take an
exposure, the command is parsed and passed to the appropriate camera
servers.  The reason for this client/server setup for the cameras is
two-fold. First, each camera requires a computer card slot and, in the
case of RAPTOR A and B, a single computer cannot hold them all.
Secondly, this setup allows us to transparently have different types
of cameras operating together on a single system.  The camera servers
all accept commands in a standard syntax and then translate them to
the hardware specific commands for that particular type of camera.


The RAPTOR analysis pipeline runs on each camera computer and
automatically processes images as they are taken.  The processing
involves subtracting a dark frame and dividing by a flat frame.
The resulting image is then processed by the SExtractor\cite{Bertin96}
program which reduces each image to a list of objects with relative
brightness and positions.  The resulting object list is then
calibrated against a catalog to get an absolute position and
brightness for each object.  Each resulting calibrated object list can
contain 10,000 to 50,000 sources depending on the instrument and the
variations in crowding for different galactic latitudes.   All of the
above operations take place within 30 seconds allowing one image to be
processed while an exposure is under way for the next image.

The calibrated object lists are then compared with calibrated object
lists from previous observations (Fig. \ref{fig:loop}).  Any object
that appears in only a single image is assumed to be non-astronomical
and discarded.  Any object which passes the above cut and shows
sufficient variability is then flagged as a possible transient.  If
both RAPTOR-A and RAPTOR-B detect a transient at the same location at
the same time a notice is sent to the alert server.  After this
occurs, all three telescopes center on the candidate and begin taking
images in rapid succession to exhaustively monitor the object.  We
believe this four-way coincidence requirement (at least two images in
two telescopes) will greatly reduce the number of false positives.

\subsection{Observatoryd}

The observatory daemon, {\sl observatoryd}, is responsible for operating the
clamshell and the weather station.  Like {\sl telescoped}, {\sl observatoryd}
accepts commands via a socket from {\sl controld}.  Only three commands are
accepted, open and close the clamshell and a status query.  The
response to the status query is to send the state of the clamshell and
the current weather conditions.  The reason for combining the
functionality of the clamshell and the weather station into a single
daemon is to allow the fastest possible reaction time for closing
the clamshell in the event of a weather alarm.  Also, in the absence
of a weather alarm, the clamshell should only move twice every night,
once to open at sunset and once to close at sunrise.  Therefore,
normal operation of {\sl observatoryd} consists almost entirely of
monitoring the weather station.

%

\section{Summary}
\label{sect:status}

The RAPTOR project is a group of robotic telescopes that will search
for optical transients.  Each system will operate in a completely
autonomous manner.  Additionally, each telescope will communicate
with the others allowing them to operate in synchronization.  Using
this technique, we will be able to eliminate false transient
detections with high efficiency.  To accomplish the task of
synchronized observations, we are developing new data acquisition
software which will coordinate the activity of the RAPTOR telescopes.
Three telescopes are currently under construction; RAPTOR A, B, and S.

In late February 2002, RAPTOR A had first light and began limited
operation in manual mode.  Since that time, construction has finished
on RAPTOR A and RAPTOR B.  We also expect construction of the RAPTOR S
telescope to be completed by the end of August 2002.  Initial testing
on all three telescopes indicate that they all will perform within
expectations.  Several scientific observations have been made with the
RAPTOR A telescope, including monitoring the eclipsing binary W
Ursae Majoris and photographing the comet Ikeya-Zhang
(Fig. \ref{fig:comet}).

RAPTOR A and B are currently able to operate in a limited robotic
mode using an early version of the RAPTOR DAQ system.  This early
version of the DAQ system allows for simple sky patrols and alert
responses.  The client/server capability of {\sl controld} is not yet
implemented, nor is the transient alert feedback loop.  However, the
automated processing pipeline is running on all of the camera
computers.  We expect the DAQ system to be fully functional in the
coming months.

\bibliography{raptordaq}
\bibliographystyle{spiebib}

\newpage

\begin{figure}
\caption{\label{fig:enclosure}The RAPTOR-B enclosure with the
clamshell open.  The weather station and power distribution panel
are visible on the side of the enclosure.}
\end{figure}

\begin{figure}
\caption{\label{fig:iobox}The RAPTOR I/O box.  The Opto 22 components
which control most of the functionality are on the bottom.}
\end{figure}

\begin{figure}
\caption{\label{fig:daq}The RAPTOR data acquisition system.  Each box
represents a specific process.  The three in bold are the primary
control processes of the system.  The arrows represent socket
connections.}
\end{figure}

\begin{figure}
\caption{\label{fig:loop}The RAPTOR transient alert feedback loop.
Each telescope must observe a transient at the same time and location
for it to be considered a valid detection.}
\end{figure}

\begin{figure}
\caption{\label{fig:comet}An 60 second exposure of comet Ikeya-Zhang
(C/2002 C1) taken with the RAPTOR-A fovea camera on 3 April 2002.}
\end{figure}

\end{document}